\documentclass{aa501}
\usepackage{graphicx}

\newcommand{\aco}{Abell~1451}
\newcommand{\rxj}{RXJ1314$-$25}

\newcommand{\cts}{\mbox{counts s$^{-1}$ arcmin$^{-2}$}}
\newcommand{\kms}{\mbox{km s$^{-1}$}}
\newcommand{\ergs}{\mbox{erg s$^{-1}$}}

\begin{document}

\title{\mbox{Abell~1451} and \mbox{1RXS~J131423.6$-$251521}: a
  multi-wavelength study of two dynamically perturbed clusters of
  galaxies}

\author{I. Valtchanov\inst{1} \and T.  Murphy\inst{2}\thanks{Now at
    Institute of Astronomy, University of Edinburgh, Royal
    Observatory, Blackford Hill, Edinburgh, EH9 3HJ} \and M.
  Pierre\inst{1} \and R. Hunstead\inst{2} \and L.  L\'emonon\inst{1}}

\institute{CEA/DSM/DAPNIA, Service d'Astrophysique, F-91191
  Gif-sur-Yvette, France \and School of
  Physics, University of Sydney, NSW 2006, Australia}

\offprints{Ivan Valtchanov, ivaltchanov@cea.fr}

\date{Received \today / Accepted \today}

\titlerunning{Abell 1451 and 1RXS J131423.6$-$251521}
\authorrunning{Valtchanov et al.}

\abstract{We present results from optical, X-ray and radio
  observations of two X-ray bright ($L_X \sim 10^{45}$ \ergs) galaxy
  clusters. \aco\ is at redshift $z=0.1989$ and has line-of-sight
  velocity dispersion $\sigma_v = 1330$ \kms\ as measured from 57
  cluster galaxies. 
  It has regular X-ray emission without signs of
  substructure, a Gaussian velocity distribution, lack of a cooling
  flow region and significant deviations from the observed scaling
  laws between luminosity, temperature and velocity dispersion,
  indicating a possible merging shock.  There is only one
  spectroscopically confirmed cluster radio galaxy, which is close to
  the X-ray peak. \\
  \mbox{1RXS~J131423.6$-$251521} (for short \rxj) has $z=0.2474$ and
  $\sigma_v = 1100$ \kms\ from 37 galaxies. 
  There are two distinct galaxy
  groups with a projected separation of $\approx 700$ kpc. The
  velocity histogram is bi-modal with a redshift-space separation of
  $\sim 1700$ \kms, and the X-ray emission is double peaked.  Although
  there are no spectroscopically confirmed cluster radio galaxies, we
  have identified a plausible relic source candidate.
   }

\maketitle

\section{Introduction}
Clusters of galaxies form a representative population which traces the
highest initial density fluctuation peaks. They are excellent tools
for exploring the distant Universe and are used to constrain
cosmological models.  A significant fraction of clusters shows
evidence of substructure (e.g. Geller \& Beers \cite{gb82}, Dressler
\& Shectman \cite{ds88}, West \cite{w94}; for a recent review see
Pierre \& Starck \cite{ps98}) and complexity in the distribution of
the different constituents --- galaxies, gas, dark matter (Baier et
al.  \cite{bai96}).  Combined multi-wavelength observations are needed
to disentangle the dynamical state of clusters.  Analysis of the
velocity and space distribution of the galaxies is very important but,
in contrast to the optical, the X-ray analysis is less prone to
projection effects and probes better the cluster mass distribution
(because the X-ray surface brightness depends on the square of the
matter density). The presence of substructure is also revealed in the
radio properties of clusters (R\"ottgering et al.  \cite{ro94}, Reid
et al.  \cite{rei98}), but the radio sources in turn may also
influence the X-ray emission (Rizza et al.  \cite{riz00}).

This paper is the third in a series of papers (Pierre et al.
\cite{mp97}, L\'emonon et al. \cite{ludo97}) dedicated to studies of
distant, bright X-ray clusters discovered in the ROSAT All-Sky Survey
(RASS, Voges et al. \cite{rass}). A sample of $\approx 10$ clusters
with $L_X > 10^{44}$ \ergs\ was selected (Pierre et al. \cite{mp94})
in the redshift range $0.1 < z < 0.3$.  In this paper we present
multi-wavelength observations of two of these clusters --- \aco\ and
\mbox{1RXS~J131423.6$-$251521} (hereafter shortened to \rxj).  General
data associated with both clusters are given in Table~\ref{tab1}.

\begin{table}

  \caption{ Properties of the two clusters.  References: Abell
    (\cite{abell}), Abell et al. (\cite{aco}) for the coordinates and
    richness; Bautz \& Morgan (\cite{bm70}) for classification for \aco
    ; coordinates for \rxj\ are from RASS (Voges et al. \cite{rass});
    $T_X$ and $L_X$ in the 2--10 keV band are from Matsumoto et al.
    (\cite{asca}); redshift and $\sigma_v$ are from this paper;  the
    apparent magnitude of an $L^{*}$ galaxy was calculated using
    $M^{*}(B_j)=-21.8$, obtained by adjusting the Lumsden et
    al. (\cite{lum97}) value of $M^{*}(B_j)=-20.16$ to the
    cosmological parameters used in this paper.}

  \label{tab1}
{\small
    \begin{tabular}{lcc}
	\hline \hline
	& \aco\ & \rxj\ \\
	\hline
    R.A. (J2000) & 12:03:16.0 & 13:14:23.6 \\
    Dec. (J2000) & $-$21:30:42 & $-$25:15:21 \\
    BM Class & III & --- \\
    Richness & 3 & --- \\
    Redshift & 0.1989 & 0.2474 \\
    $\sigma_v$ km~s$^{-1}$ & 1330 & 1100 \\
    $B_j^{*}$ & 19.5 & 20.2 \\
    $L_X\ (10^{45})$ erg~s$^{-1}$ & 1.5 & 1.8 \\
    $T_X$ keV & 13.4 & 8.7 \\
    \hline
\end{tabular} }
\end{table}

The plan of the paper is as follows: in Section~\ref{sec:opt} we
present optical observations, data reduction, redshift catalogues and
data analysis for the two clusters. In Section~\ref{sec:x} and
\ref{sec:radio} we present ROSAT-HRI observations and data analysis
and ATCA radio observations and data analysis respectively.  Finally
in Section~\ref{sec:disc} we discuss the multi-wavelength view of
\aco\ and \rxj. Throughout the paper we use $H_0 = 50$ \kms\  Mpc$^{-1}$
and $q_0 = 0.5$.

\section{Optical observations, data reduction and analysis}

\label{sec:opt}

\subsection{Observations and reduction}

\begin{figure*}
  \includegraphics[width=16cm]{f1.bb}
  \caption{\aco: V-band image from the Danish 1.5-m telescope showing
    the objects included in the spectroscopic study. Numbers correspond
    to the object identification in Table~\ref{tab3}; those denoted as
    diamonds are either stars or non-cluster members. The dashed box is
    the central part, shown zoomed in Fig.~\ref{fig2}.}

  \label{fig1}
\end{figure*}

Multi-Object Slit (MOS) spectra for both clusters were acquired using
the ESO 3.6-m telescope at La Silla in two runs, one in 1993 and one in
1999 (Table~\ref{tab2}). The first run was equipped with EFOSC1 (ESO
Faint Object Spectrograph and Camera), Tektronix 512$\times$512 CCD
(pixel size 27 $\mu$m and $0.61\arcsec$/pixel) and Grism B300, covering
the   range 3740--6950 \AA, with central wavelength 5250 \AA\ and
dispersion of 6.2\AA/pixel. The observations in 1999 were performed
with EFOSC2, CCD \#40 2048$\times$2048 (pixel size 15 $\mu$m,
$0.157\arcsec$/pixel) and Grism O300 which covers 3860--8070\AA\, with
central wavelength 5000 \AA\ and dispersion 2.06 \AA/pixel. However,
the final wavelength range depends on the position of each slit on the
mask. The seeing was $1.0-1.2\arcsec$ for both 1993 and 1999 runs.
The corresponding airmasses and the exposure times for each mask/slit
configuration are given in Table~\ref{tab2}.

\begin{figure}[htb]
 \centering
  \includegraphics[height=8cm]{f2.bb}

  \caption{\aco: zoomed image of the central part of Fig.~\ref{fig1}.
    The plus sign marks the X-ray emission centre (see
    Section~\ref{sec:x}) and the contours are ATCA 13cm radio
    observation with levels = 0.3, 0.5, 1, 2, 5 and 10 mJy/beam,  rms
    noise level is $60\mu$Jy/beam (see Section~\ref{sec:radio}). The
    object in brackets is a background  galaxy.}

  \label{fig2}
\end{figure}

\begin{table*}

\caption{Summary of the multi-slit spectroscopic observations} 

\label{tab2}
{\scriptsize
\begin{tabular}{ll@{\hspace{2mm}}c@{\hspace{2mm}}ccc}
\hline
\hline
\multicolumn{1}{c}{Date} &  \multicolumn{1}{c}{Object} &
\multicolumn{1}{c}{RA \hspace{7mm} Dec} & Airmass & \multicolumn{1}{c}{MOS} 
& \multicolumn{1}{c}{Exposure}\\
 &  & \multicolumn{1}{c}{(J2000)} &  & masks/slits & 
\multicolumn{1}{c}{(s)}\\

\hline
1993 Mar 29   & \aco    & 12:03:15 $-$21:31:36 & 1.2 & 1/15 &  900\\
1993 Mar 29   & \rxj    & 13:14:29 $-$25:16:25 & 1.1 & 1/15 &  2x900+1800\\[8pt]
1999 Apr 19/20& \aco E & 12:03:12 $-$21:33:25 & 1.34 & 2/20+21 &  3x1800/4x2100\\
1999 Apr 19/20& \aco W & 12:03:23 $-$21:33:25 & 1.40 & 2/21+19 &  2x3600/4200\\
1999 Apr 19/20& \rxj N & 13:14:22 $-$25:14:55 & 1.06 & 2/17+18 &  2x3600/4x2100\\
1999 Apr 19/20& \rxj S & 13:14:22 $-$25:17:55 & 1.04 & 2/17+18 &  4x1800/4x2100\\
\hline
\end{tabular}
}
\end{table*}

All spectra were processed within the ESO-MIDAS or IRAF environment to
produce the final wavelength calibrated, sky subtracted spectra. A
HeAr lamp was used for wavelength calibration, and a
second-order cubic spline or polynomial fit to 5--12
comparison lines gave rms residuals $<$1\AA.

We have determined redshifts using the Tonry \& Davis (\cite{TD79})
cross-correlation technique as implemented in the RVSAO package (Kurtz
\& Mink \cite{rvsao}). Before processing with RVSAO we have masked out
spectral ranges where significant residuals from sky subtraction could
occur, and also any possible emission lines. Generally we have
cross-correlated our observed spectra with 3 to 6 galaxy and star
templates with good signal-to-noise, and also with the composite
absorption-line template (fabtemp97) distributed with RVSAO. For each
target object we adopted the redshift value from the best (highest
correlation coefficient) template. The redshifts were then checked by
identifying the prominent absorption features (CaII H \& K, G band,
MgI triplet) in each spectrum.  If emission lines were present in the
spectra, redshifts were determined from them.

The errors in redshift were computed from $\Delta z = k/(1+r)$, where
$r$ is the cross-correlation coefficient and $k$ was determined
empirically by adding noise to a high signal-to-noise spectrum and
correlating it with template spectra with known velocities.  For our
observational configurations we found $k=0.003$, which gives a velocity
error of about 200 \kms\  for a redshift estimate with $r=3$. We have
checked our redshift measurements and uncertainties using a few galaxy
spectra repeated in different runs or masks and they are all
in excellent agreement, consistent with the derived errors.

We have not converted our measurements to the heliocentric system
because the correction is about $1$ \kms\ for the 1993 run and $-8.8$
\kms\ for the 1999 run, well below the uncertainties.

The final redshift catalogue of objects for \aco\ is shown in
Table~\ref{tab3} and for \rxj\ in Table~\ref{tab4}, with corresponding
finding charts in Figs~\ref{fig1} and \ref{fig3}.

\begin{figure*}
  \includegraphics[width={\hsize}]{f3.bb}
  \caption{\rxj: I-band image from the Danish 1.5-m telescope showing
    the objects in the spectroscopic study. Numbers correspond to the
    object identifications in Table~\ref{tab4}; those marked with diamonds
    denote either stars or non-cluster members. Note that we have not
    plotted galaxy \#20 because it is not resolved spatially
    from galaxy \#19.}
  \label{fig3}
\end{figure*}

\subsection{Optical data analysis}
\subsubsection{\aco}

\begin{figure}
 \centering
  \includegraphics[width=8cm]{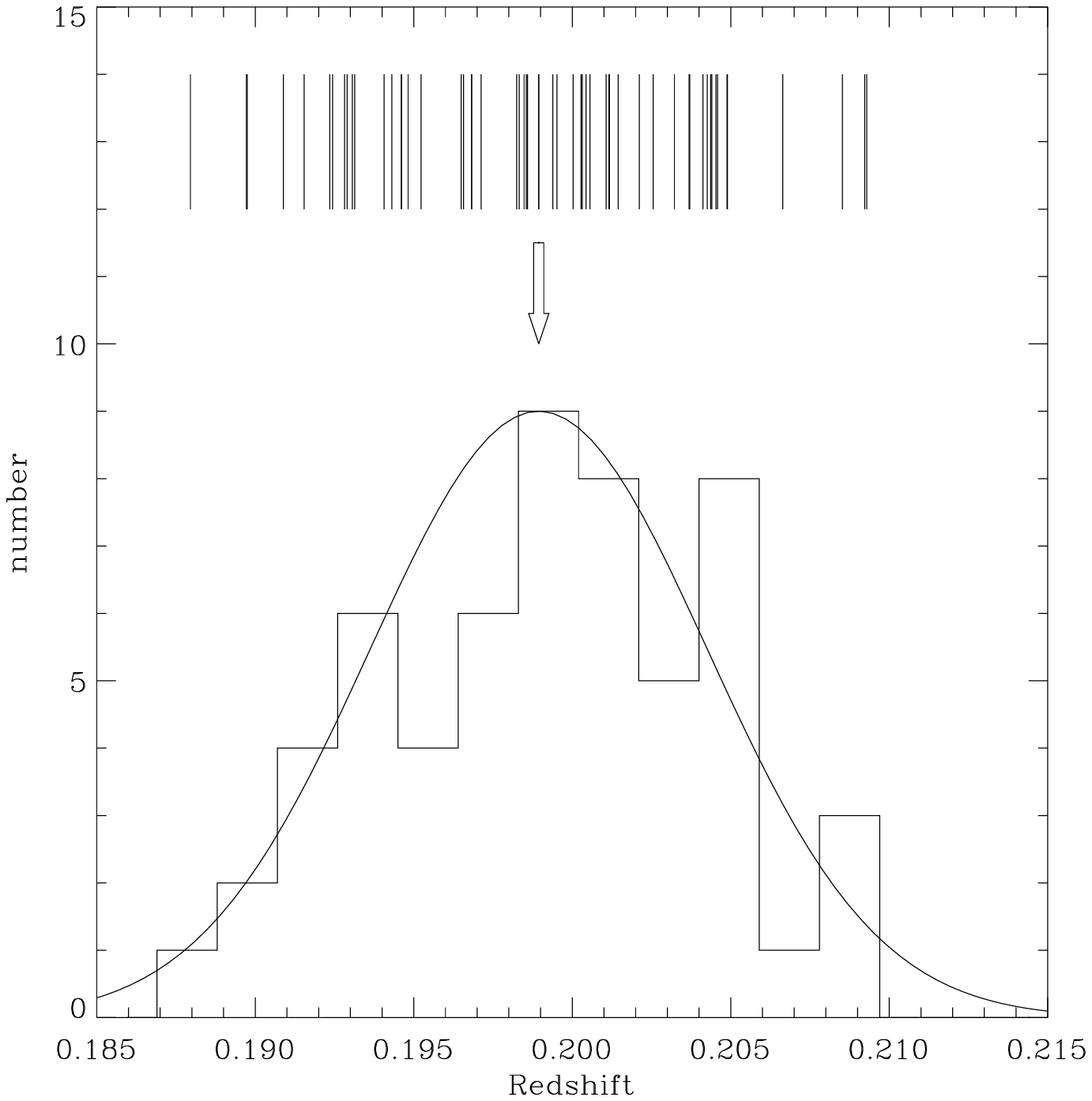}
  \caption{\aco:
    Redshift histogram for cluster members; the bin size is 0.0019.
    The bi-weighted location ($C_{BI}$) is shown by an arrow. The
    lines above the histogram are the actual \mbox{1-D} redshift
    distribution and the continuous line is the Gaussian function with
    bi-weighted location ($C_{BI}$) and bi-weighted scale ($S_{BI}$)
    from Table~\ref{tab4}.  }
  \label{fig4}
\end{figure}

Cluster membership was determined by using an interactive version of
the Beers et al. (\cite{rostat}) ROSTAT package (Valtchanov
\cite{val99}) and confirmed by the 3-$\sigma$ clipping method of Yahil
\& Vidal (\cite{yv77}).  The cluster redshift distribution is shown in
Fig.~\ref{fig4} and some relevant characteristics, taking
redshift errors into account, are given in Table~\ref{tab5}. The
distribution is very close to Gaussian, which has been
quantified by various statistical tests for normality (D'Agostino \&
Stephens \cite{ds86}) --- the Wilk-Shapiro test accepts normality at
the 99\% level and the Anderson-Darling test at the 95\% level.

\setcounter{table}{4}
\begin{table}
  \caption{\aco: Some statistical characteristics of the redshift
    distribution, corrected for measurement errors (Danese et al.
    \cite{d80}). The scale measure is in the cluster rest frame, i.e.,
    $S_{BI} = c\sigma_z/(1+z)$ (Harrison \cite{h74}). The errors
    ($1\sigma$) are calculated by using the accelerated bias-corrected
    bootstrap technique with 1000 bootstrap resamplings (Efron \&
    Tibshirani \cite{et86})}
  \label{tab5}
{\small
\begin{tabular}{ll}
\hline
\hline
\multicolumn{1}{c}{Characteristic} & \multicolumn{1}{c}{Value} \\
\hline
\multicolumn{2}{c}{N = 57}\\[2mm]
Bi-weighted location: $C_{BI}$ &  $\overline{z}=0.1989^{+0.0005}_{-0.0007}$\\
Bi-weighted scale: $S_{BI}$ &  $1330^{+130}_{-90}$ \kms\ \\
Maximum gap                &    564 \kms\ \\
\hline
\end{tabular}
}
\end{table}

There is no single dominant cluster galaxy. The brightest cluster
members (BCMs, \#34, \#39, \#36 and \#40) are linearly aligned in
projection (see Fig.~\ref{fig2}) and their spectra are typical
elliptical galaxy absorption spectra with no emission lines.  The
projected galaxy density distribution from SuperCOSMOS (Hambly et al.
\cite{sss}) data shows an elongation in the same direction. The
velocity dispersion among the BCMs is $\sim 800$ \kms, somewhat less
than the overall cluster velocity dispersion.

\subsubsection{\rxj}
\label{rxj:optic}
\begin{figure}
 \centering
  \includegraphics[width=8cm]{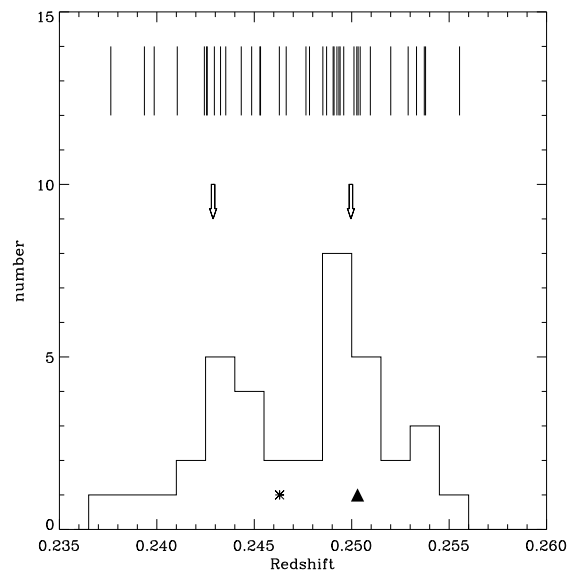}
  \caption{\rxj:
    Redshift histogram for cluster members; the bin size is 0.0015.
    The $C_{BI}$ locations for the two groups (see Table~\ref{tab6})
    are indicated by arrows. The two brightest cluster galaxies are
    marked: \#48 with an asterisk and \#19 with a filled triangle.}
  \label{fig5}
\end{figure}

\begin{figure}[h]
 \centering
  \includegraphics[width=8cm]{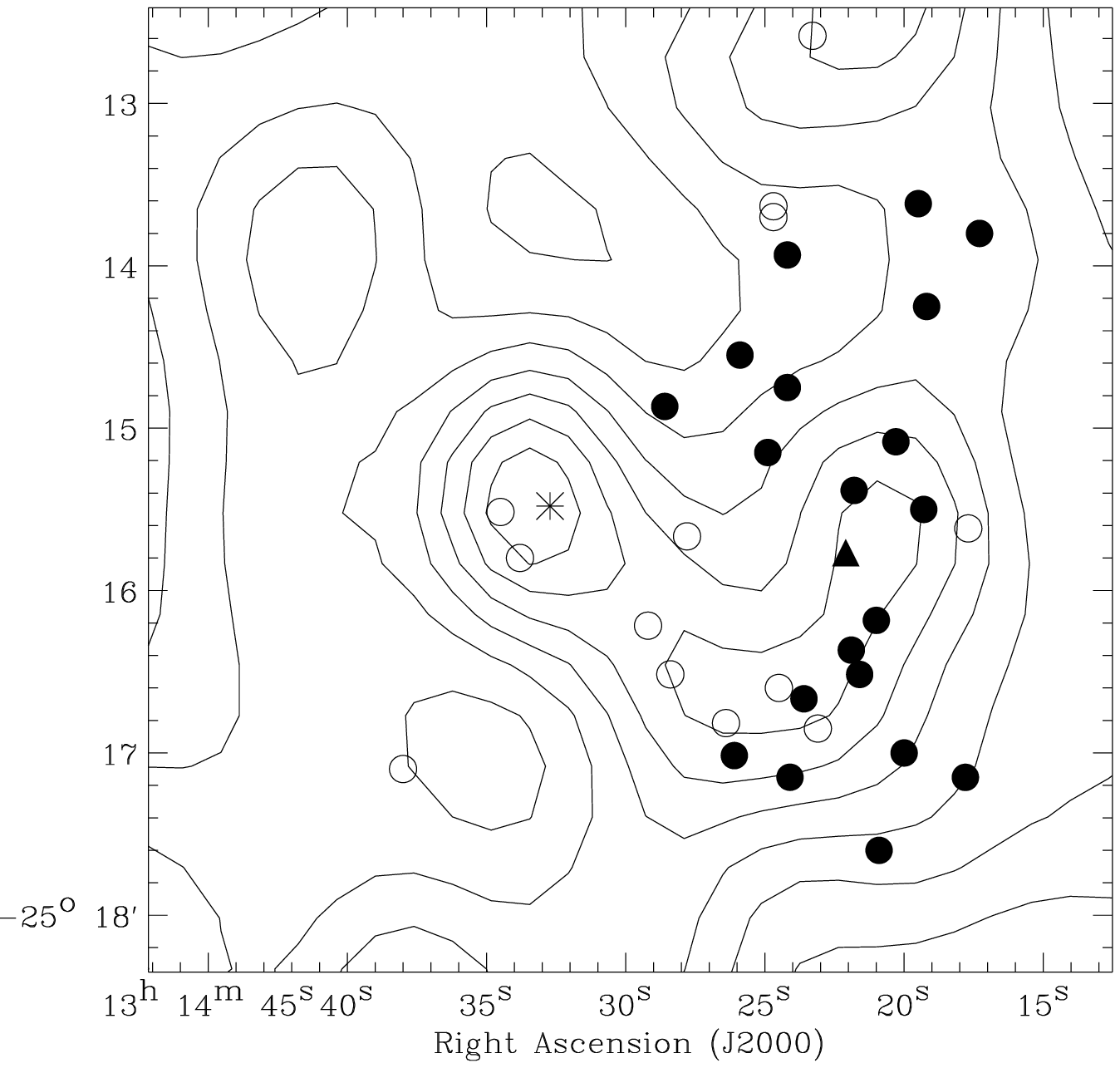}
  \caption{\rxj: Sky distribution for cluster members. Open circles
    denote the members of the eastern group, while the filled circles
    are those belonging to the western group as assigned by KMM (see
    Table~\ref{tab6}). The contours are the adaptive kernel density
    estimate (Silverman \cite{sil86}, Pisani \cite{p96}) of the
    SuperCOSMOS galaxy distribution. The asterisk and filled triangle
    mark the positions of the first- and second-ranked BCGs
    respectively (\#48 and \#19 in Table~\ref{tab4}). Note that galaxy
    \#20 is too close to galaxy \#19 to be plotted separately.}
  \label{fig6}
\end{figure}

Statistics of the redshift distribution are shown in Table~\ref{tab6}
and the redshift histogram in Fig.~\ref{fig5}. The distribution
clearly shows two peaks and the test for the hypothesis of a unimodal
distribution --- the dip test (Hartigan \cite{h85}) --- gives an
insignificant probability of a single mode.  Going further, we have
applied the KMM method for detecting bimodality (Ashman et al.
\cite{kmm}), which confirmed the dip test negative result for
unimodality and also gave us the probable group membership, assuming a
bi-modal distribution.  In Table~\ref{tab6} we show the same
statistics for both groups.  Although the number of cluster members
is relatively small for giving high weight to the KMM
result, the segregation of the galaxies on the sky supports that
conclusion (see Fig.~\ref{fig6}).

\begin{table}
\caption{\rxj: Statistics of the redshift distribution, as for
Table~\ref{tab5}.  The subdivision into two groups follows from KMM
(Ashman et al. \cite{kmm})}
\label{tab6}
{\small
\begin{tabular}{ll}
\hline
\hline
\multicolumn{1}{c}{Characteristic} & \multicolumn{1}{c}{Value} \\
\hline
\multicolumn{2}{c}{Total, N = 37}\\[2mm]
$C_{BI}$ &  $\overline{z}=0.2474^{+0.0006}_{-0.0008}$\\
$S_{BI}$ &   $1100^{+140}_{-90}$ \kms\ \\
Maximum gap                &    523 \kms\ \\
\hline
\multicolumn{2}{c}{Group 1 (East), N = 15}\\[2mm]
$C_{BI}$ &  $\overline{z}=0.2429^{+0.0003}_{-0.0008}$\\
$S_{BI}$ &   $590^{+110}_{-150}$ \kms\ \\
Maximum gap                &    516 \kms\ \\
\hline
\multicolumn{2}{c}{Group 2 (West), N = 22}\\[2mm]
$C_{BI}$ &  $\overline{z}=0.2500^{+0.0006}_{-0.0005}$\\
$S_{BI}$ &   $560^{+120}_{-70}$ \kms\ \\
Maximum gap                &    523 \kms\ \\
\hline
\end{tabular}
}
\end{table}

The optical image of the cluster (Fig.~\ref{fig3}) shows two dominant
galaxies (\#48 and \#19, marked also in Table~\ref{tab4}) as the
brightest cluster galaxies (BCG). Both have typical giant elliptical
galaxy absorption line spectra, without any apparent emission lines.
The second-ranked BCG (\#19, $z=0.2503$) is at rest with respect to
the galaxy members of the western group, while the first-ranked galaxy
(\#48, $z=0.2463$) differs by $\sim 800$ \kms\ from the mean redshift 
of the eastern group, but lies closer to the overall cluster mean. The
separation between the two groups in velocity space is $\sim 1700$
\kms, substantially greater than their combined velocity dispersions.
With optical data alone we can not determine whether the two groups
are gravitationally connected.

\section{X-ray data and analysis}
\label{sec:x}

The X-ray observing logs for both clusters are given in
Table~\ref{tab7}.  For both ROSAT/HRI images we have used
$5\arcsec$/pixel binning of the event list in order to reduce the noise
as much as possible without losing information on the cluster
extension. The raw photon images were then filtered using wavelet
analysis with Poisson noise modeling (Starck \& Pierre \cite{sp98}) at
$10^{-4}\ (\sim 4\sigma)$ significance level for the wavelet
coefficients.

\begin{table}
\caption{ROSAT-HRI X-ray observing log}
\begin{tabular}{lcc}
\hline
\hline
\multicolumn{1}{c}{Cluster} & Date & Exposure (s) \\
\hline
\aco & 1997 Jul 14--16 & 25603 \\
\rxj & 1996 Jan 27--31 & 29294 \\
\hline
\end{tabular}
\label{tab7}
\end{table}

\subsection{X-ray morphology}

The X-ray contours are shown overlaid on optical images and radio
observations (see Sec.\ref{sec:radio}) in Fig.~\ref{fig7} for \aco\ 
and in Fig.~\ref{fig9} for \rxj.

\begin{figure*}
 \centering
  \includegraphics[width=18cm]{f7.bb}
  \caption{\aco: Optical V-band image from the Danish 1.5-m telescope
    overlaid with HRI X-ray contours (heavy) and 20 cm ATCA radio contours 
    (lightweight). The X-ray contours run from 0.006 to 0.02
    \cts\ in logarithmic steps; the innermost cluster contour is at
    0.0134 \cts. The peak of the extended X-ray emission, used for the
    profile measurements, is marked with a cross. The radio contour
    levels are 0.2, 0.3, 0.5, 1, 2, 5 and 10 mJy/beam; the rms
    noise level is 80 $\mu$Jy/beam.}
  \label{fig7}
\end{figure*}

\begin{figure*}
 \centering
  \includegraphics[width=18cm]{f8.bb}
  \caption{\rxj: Optical I-band image from the Danish 1.5-m telescope
    overlaid with HRI X-ray contours (heavy) and 20 cm ATCA contours 
    (lightweight). The X-ray contours run from 0.004 to 0.065
    \cts\ in logarithmic steps; the innermost cluster contour is at
    0.012 \cts.  The centre of the X-ray emission taken for the
    profile measurements is marked with a cross. The radio contour
    levels are 0.35, 0.5, 0.7, 1, 1.5 and 2 mJy/beam; the rms noise
    level is 90 $\mu$Jy/beam.}
  \label{fig9}
\end{figure*}

\subsubsection{\aco}
The X-ray image in Fig.~\ref{fig7} shows a very strong point source 
$\sim 1\arcmin30\arcsec$ south of the cluster centre which coincides
with a QSO at $z=1.17$ (object \#32 in Table~\ref{tab3}). The cluster
emission is regular, with the inner contours slightly twisted but no
sign of substructure.  The X-ray emission peaks at R.A. =
$12^h03^m16\fs6$ and Dec. = $-21\degr32\arcmin21\arcsec$, which is 
$36\arcsec$ (150 kpc) north of the brightest cluster galaxy (\#34 in
Table~\ref{tab3}) and $12\arcsec$ west of the second brightest galaxy
(\#39), and $\sim 1\arcmin40\arcsec$ from the catalogued cluster
position (Abell et al. \cite{aco}).

\subsubsection{\rxj}
The cluster X-ray emission is quite irregular, showing two central
peaks and a SE--NW elongation.  Unfortunately, there is an X-ray
emitting star projected in front of the cluster (object \#29 in
Table~\ref{tab4}) only $\sim 24\arcsec$ from the adopted X-ray centre.
The strong point source SE of the cluster centre is a Sy1 galaxy
(object \#47 in Table~\ref{tab4}) and a cluster member. There is no
indication of substructure in the X-ray emission associated with the
eastern group or the brightest cluster galaxy (see Fig.~\ref{fig6}).

\subsection{X-ray properties}

\begin{figure}
 \centering
  \includegraphics[height=8cm,angle=270]{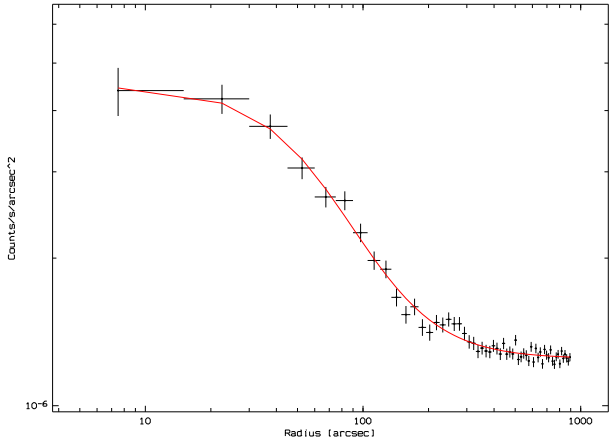}
  \caption{\aco:
    Circularly averaged surface brightness profile in the energy band
    [0.1--2.4] keV. Bin size is $15\arcsec$ and the solid line is the
    King profile fit (Eq.~\ref{king}) with parameters in
    Table~\ref{tab:king}.}
  \label{fig11}
\end{figure}

\begin{figure}
 \centering
  \includegraphics[height=8cm,angle=270]{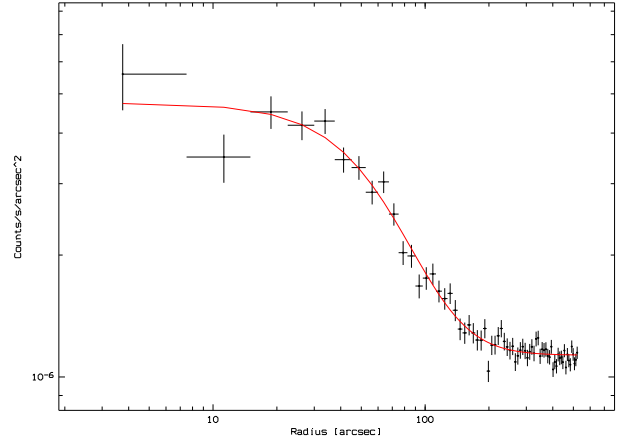}
  \caption{\rxj:
    Circularly averaged surface brightness profile in the [0.1--2.4] keV
    energy band. Bin size is $7.5\arcsec$ and the solid line is the
    King profile fit (Eq.~\ref{king}). The deviation of the inner two
    points from the model is due to the irregular morphology in the
    cluster centre.}
  \label{fig12}
\end{figure}

To estimate the basic physical cluster parameters we model the X-ray
surface brightness using the isothermal $\beta$-model (King
\cite{king62}, Cavaliere \& Fusco-Femiano \cite{cff76})

\begin{equation}
S(r) = S_0\left(1 + (r/r_c)^2\right)^{0.5 - 3\beta} + S_b,
\label{king}
\end{equation}
where $S(r)$ is the azimuthally averaged X-ray surface brightness as a
function of the radial distance $r$ from the centre, $S_0$ is the
central brightness, $S_b$ is the background contribution and $r_c$ is
the core radius. To fit the model profile to the data we define a
proper cluster centre and exclude any discrete sources projected over
the cluster X-ray emission. Finally, we obtain an average profile by
summing the cluster X-ray photons in concentric rings and fit the
model (Eq.~\ref{king}) with $S_0,\ S_b,\ \beta$ and $r_c$ as free
parameters, taking into account the Poissonian errors of the rings
counts.  The X-ray surface brightness profiles of \aco\ and \rxj\ are
shown in Figs.~\ref{fig11}--\ref{fig12}, with the corresponding
parameters in Table~\ref{tab:king}.

\begin{table}
\caption{$\beta$-profile best-fit parameters with the corresponding
  95\% confidence intervals.}
\label{tab:king}
{\small
\begin{tabular}{lcc}
\hline
\hline
\multicolumn{1}{c}{Parameter} & \multicolumn{1}{c}{\aco} &
\multicolumn{1}{c}{\rxj} \\
\hline
$r_c$ (arcsec)   & $59\pm20$ &  $81\pm 28$\\
$r_c$ (kpc)      & $240\pm90$   &  $400\pm140$ \\
$\beta$ & $0.50\pm0.08$ & $0.77\pm0.23$ \\
\hline
\end{tabular}
}
\end{table}

To derive the count-rate in the [0.1--2.4] keV ROSAT/HRI band we
integrate the fitted surface brightness profile analytically,
excluding the background. The integration is usually carried out to a
given radius $r_{\rm lim}$, where the surface brightness profile
reaches the detection limit; for both clusters we put $r_{\rm
lim}=300\arcsec$. For the overall count rate $C$ we have
\begin{eqnarray}
\label{x:counts}
C(<r_{\rm lim}) & = &  \int_{0}^{r_{\rm lim}} 2\pi r S(r)dr =  \frac{\pi S_0
  r_c^2}{3/2-3\beta} \times \\ \nonumber
 & & \times \left[\left
    (1+\left(r_{\rm lim}/r_c\right)^2\right)^{3/2-3\beta} - 1\right].
\end{eqnarray}
By means of EXSAS (Zimmermann et al. \cite{exsas}), the observed
count-rate is then used to normalize the spectral model --- a Raymond
\& Smith (\cite{rs77}) thermal plasma emission, convolved by the
ROSAT/HRI response function, with temperature and metallicity from
ASCA data (Matsumoto et al.\ \cite{asca}) and line-of-sight Galactic
absorption by H\,I from Dickey \& Lockman (\cite{dl90}). The
spectrum is then integrated to derive the flux and cluster rest frame
luminosity in the ``standard'' X-ray bands [0.5--2] and [2--10] keV.
The profile model, together with the emission model, was used to
derive the emission measure, $\int_{V} n_e n_p dV$, and the proton
density distribution:
\begin{equation}
\label{eq:np}
n_p(r) = n_p(0)\left(1 + (r/r_c)^2\right)^{-3\beta/2}.
\end{equation}
Assuming a hydrogen gas then $\rho_{\rm gas} = 2.21 \mu m_p n_p$ and
we can calculate the mass of the gas inside a given radius $R$:
\begin{equation}
M_{\rm gas}(r<R) = \int_0^R 4\pi \rho_{\rm gas}(r) dr.
\end{equation}

The total gravitating mass of the cluster within radius $r$ can be
estimated, assuming hydrostatic equilibrium, as 
\begin{equation}
M_{\rm tot}(r) = \frac{r^2 kT}{G\mu m_p}\left(\frac{1}{n}
  \frac{dn}{dr} + \frac{1}{T}\frac{dT}{dr}\right),
\end{equation}
with $n = n_e + n_p = 2.21 n_p$ and $\mu = 0.61$ for the abundances.
Ignoring any radial temperature dependence we can derive:
\begin{equation}
M_{\rm tot}(r) = 3\beta r^3kT/G\mu m_p (r^2+r_c^2).
\end{equation}

The derived and model parameters for the X-ray emission of both
clusters are shown in Table~\ref{tab:x}. There are small differences in
luminosity when compared with ASCA data. This can be explained,
firstly, by the smaller limiting distance ($4\arcmin$) for flux
integration adopted in Matsumoto et al. (\cite{asca}) and secondly, by
the fact that for both clusters there are discrete X-ray sources within
the extended cluster emission which were not resolved by ASCA and so
their contribution was not subtracted.

Derived cluster masses should be considered with caution, as the
combined X-ray/optical analysis tends to indicate that neither cluster
has reached equilibrium.

Although the X-ray emission in both clusters is not centrally peaked,
we have tried to estimate the cooling flow radius --- the zone where
the time for isobaric cooling is less than the age of the universe
(Sarazin \cite{s86}, Fabian \cite{f94}). For any reasonable choice of
the Hubble constant and gas parameters ($n_p,\ T_X,\ Z$) there is no
such zone.

\begin{table}
\caption{X-ray data for both clusters. $N_H$ is taken from Dickey and
  Lockman (\cite{dl90}). ASCA data for $T_X$ and $L_X$ in the $[2-10]$
  keV band (Matsumoto et al. \cite{asca}) are indicated
  correspondingly.  The count-rate $C$ is in [0.1--2.4] keV band. The
  luminosities in [0.5--2], [2--10] keV and bolometric bands were
  obtained by extrapolation of the King profile out to $5\arcmin$.}
\label{tab:x}
{\scriptsize
\begin{tabular}{lcc}
\hline
\hline
\multicolumn{1}{c}{Parameter} & \multicolumn{1}{c}{\aco} &
\multicolumn{1}{c}{\rxj} \\
\hline
\\[-2mm]
$r_{\rm lim} = 5\arcmin$ (Mpc) & 1.3 & 1.5 \\
$N_H$ ($10^{20}$ cm$^{-2}$) & 4.5 & 6.7 \\
$T_X$ (keV) ASCA & $13.4^{+1.9}_{-1.5}$ & $8.7^{+0.7}_{-0.6}$ \\
Count-rate $C(r < r_{\rm lim})$ (cts s$^{-1}$) & 0.126 & 0.083 \\

$F_X\ [$0.5--2$]$ keV ($10^{-12}$ erg\,cm$^{-2}$\,s$^{-1}$) & 4.4 & 3.2 \\
$F_X\ [$2--10$]$ keV ($10^{-12}$ erg\,cm$^{-2}$\,s$^{-1}$) & 11.2 & 6.8 \\

$L_X\ [$0.5--2$]$ keV ($10^{44}$ erg s$^{-1}$) & 6.8 & 7.4 \\
$L_X\ [$2--10$]$ keV ($10^{44}$ erg s$^{-1}$) & 17.2 & 16.0 \\
$L_X\ [$2--10$]$ keV ($10^{44}$ erg s$^{-1}$) ASCA & 15.0{\rlap{$^\dag$}} & 18.0{\rlap{$^\ddag$}} \\
$L_{bol}\ (10^{44}$ erg s$^{-1}$) & 39.8 & 34.0 \\

$n_p(0)$ (10$^{-3}$ cm$^{-3}$) &  5.09 & 5.31 \\

$M_{\rm tot}(<r_{\rm lim})$ (10$^{14} {\rm M}_{\sun}$)$^{*}$ & 8.6 & 9.7\\
$M_{\rm gas}(<r_{\rm lim})$ (10$^{14} {\rm M}_{\sun}$)$^{*}$ & 2.2 & 2.6\\
$M_{\rm gas}/M_{\rm tot}(< r_{\rm lim})$ & 0.25 & 0.27 \\
\\[-2mm]
\hline
\\[-2mm]
\end{tabular}}

{\footnotesize $^\dag$ ASCA/GIS luminosity out to $4\arcmin$, including 
the contribution from the QSO $\sim 1\arcmin 30\arcsec$ south of the centre.\\
$^\ddag$ ASCA/GIS luminosity out to $4\arcmin$, including
the contribution from the Sy1 galaxy $\sim 3\arcmin 15\arcsec$
south-east of the centre. \\
$^{*}$ Assuming hydrostatic equilibrium.
}
\end{table}

\section{Radio observations and data analysis}
\label{sec:radio}

\subsection{Observations and reduction}

The Australia Telescope Compact Array (ATCA) consists of five 22 m
antennas on a 3 km east-west railway track, and a sixth antenna 3 km
from the western end of the track.  Each cluster was observed in 1999
February for 12 hours with the ATCA in the 6C configuration, giving
baselines ranging from 153 m to 6 km. Simultaneous observations at
1.384 GHz and 2.496 GHz were made for each cluster.

The radio data were processed using standard {\tt MIRIAD} (Sault et
al.  \cite{saul95}) software and techniques. The primary flux density
calibrator was PKS B1934$-$638, with PKS B1245$-$197 as the secondary
phase calibrator.  The data were then CLEANed and
RESTORed. Table~\ref{beams} lists details of the ATCA observations,
including the size and orientation of the elliptical Gaussian
restoring beams and the rms noise in the final images.

\begin{table}
\caption{\label{beams} Details of the ATCA radio observations of the
two clusters. }
{\scriptsize
\begin{tabular}{cccccr}\hline
                  &           &           &  \multicolumn{3}{c}{Restoring Beam} \\
  Cluster         & Frequency & RMS Noise & $b_{\rm maj}$ & $b_{\rm min}$ & PA \\
  (Obs. Date)          &  (GHz)    & mJy/beam         & ($''$)& ($''$)& (deg) \\ \hline\hline\\[-2mm]
\aco              & 1.384      &   0.08    & 27.8 & 9.4 & $-$0.5 \\
(1999 Feb 25)     & 2.496      &   0.06    & 15.4 & 5.2 & $-$0.5 \\[3mm]
\rxj              & 1.384      &   0.09    & 23.8 & 9.4 & 0.9 \\
(1999 Feb 26)     & 2.496      &   0.06    & 13.2 & 5.2 & 0.8 \\ \hline
\end{tabular}
}
\end{table}

\subsection{Data Analysis}

The {\tt AIPS} task {\tt VSAD} was used to generate a radio source list
for each cluster, above a nominal cutoff of 0.5 mJy at 1384 MHz. An
elliptical Gaussian was fitted to each source as described by Condon
(\cite{cond97}). For sources which were extended or complex, the
integrated flux density was estimated by using {\tt kview} (Gooch
\cite{goo96}) to sum inside a rectangular region defined around the
source. The radio sources for \aco\ and \rxj\ within one Abell radius
($R_A = 1.7/z$ arcmin) of the cluster centre\footnote{For \aco\ we have
adopted the X-ray peak as the centre of the cluster.}  are listed in
Table~\ref{radiosrcs}.  Positions are measured from the 1384 MHz image
except for sources R1 and R9 in \aco\ which are measured at 2496 MHz;
quoted errors are the quadratic combination of the formal {\tt VSAD}
error and a nominal 0.5\arcsec\ calibration uncertainty.  Errors in
flux density are taken directly from the {\tt VSAD} output; if no error
is quoted for $S_{\rm int}$ the measurement was made using {\tt kview}.

\begin{figure*} 
      \includegraphics[width=18cm]{f11.bb}
  \caption{\aco:
    DSS image overlaid with ATCA 20cm radio contours.  The Abell
    radius with X-ray emission peak as a cluster centre is shown as a
    green circle. The objects are designated by their respective names
    from Table~\ref{radiosrcs}.  The radio contour levels are
    0.2, 0.3, 0.5, 1, 2, 5 and 10 mJy/beam; the rms noise level is
    80 $\mu$Jy/beam.{\bf [Electronic version only]}}
  \label{fig13}
\end{figure*}

\begin{figure*} 
      \includegraphics[width=18cm]{f12.bb}
  \caption{\rxj:
    DSS image overlaid with ATCA 20cm radio contours.  The Abell
    radius is shown as a green circle. The objects are designated by
    their respective names from Table~\ref{radiosrcs}.  The radio
    contour levels are 0.35, 0.5, 0.7, 1, 1.5 and 2 mJy/beam; the rms
    noise level is 90 $\mu$Jy/beam.{\bf [Electronic version only]}}
  \label{fig14}
\end{figure*}

\setlength{\tabcolsep}{0.1cm}

\begin{table}
\caption{Radio sources for \aco\ and \rxj\ within 1 Abell radius of the
  cluster centre. Source coordinates are from {\tt VSAD}. The peak
  ($S_{\rm peak}$) and integrated ($S_{\rm int}$) flux  densities were
  measured using either {\tt VSAD} or {\tt kview} as
  described in the text. For \aco\ we adopt the X-ray emission peak as
  the centre.}
{\scriptsize
  \begin{tabular}{rcccll}
\hline
    \hline
    N  & RA (err s) & Dec (err $\arcsec$) & $\nu$ &
    $S_{\rm peak}$(err) & $S_{\rm int}$(err) \\
    &\multicolumn{2}{c}{(J2000)} & GHz   & mJy/bm & mJy \\
    \hline \\
    \multicolumn{6}{c}{\normalsize \aco} \\[2mm]
R1  & 12:02:51.67(0.04) & $-$21:26:35.8(1.0) & 1.384 & 3.9        & 12.1 \\
    &                   &                    & 2.496 & 1.6        & 9.6  \\
R2  & 12:02:56.73(0.06) & $-$21:28:46.7(1.9) & 1.384 & 0.60(0.06) & 1.10(0.19) \\
   &                   &                     & 2.496 & ---        & --- \\
R3  & 12:02:58.94(0.04) & $-$21:38:35.8(0.5) & 1.384 & 25.8(0.08) & 28.3(0.15) \\
   &                   &                     & 2.496 & 12.2(0.06) & 14.4(0.12) \\
R4  & 12:03:06.66(0.05) & $-$21:39:29.8(0.5) & 1.384 & 18.6(0.08) & 18.6(0.14) \\
   &                   &                     & 2.496 & 7.80(0.06) & 8.27(0.11) \\
R5  & 12:03:08.65(0.05) & $-$21:39:40.6(0.6) & 1.384 & 3.75(0.08) & 4.24(0.15) \\
   &                   &                     & 2.496 & 1.95(0.06) & 2.31(0.12) \\
R6  & 12:03:10.59(0.04) & $-$21:29:54.4(0.7) & 1.384 & 1.90(0.08) & 1.97(0.14) \\
   &                   &                     & 2.496 & 0.98(0.06) & 1.02(0.11) \\
R7  & 12:03:17.35(0.04) & $-$21:32:31.3(0.5) & 1.384 & 11.4(0.07) & 15.4  \\
   &                   &                     & 2.496 & 5.51(0.06) & 7.2 \\
R8  & 12:03:26.97(0.04) & $-$21:30:49.9(0.7) & 1.384 & 2.21(0.07) & 2.65(0.15) \\
   &                   &                     & 2.496 & 1.19(0.06) & 1.36(0.11) \\
R9  & 12:03:32.53(0.04) & $-$21:33:09.1(0.5) & 1.384 & 6.11(0.07) & 9.7 \\
   &                   &                     & 2.496 & 2.56(0.04) & 6.2  \\
R10 & 12:03:32.85(0.05) & $-$21:36:26.2(1.6) & 1.384 & 0.73(0.08) & 0.73(0.10) \\
   &                   &                     & 2.496 & 0.36(0.06) & 0.40(0.11) \\
R11 & 12:03:33.84(0.05) & $-$21:30:22.1(1.7) & 1.384 & 0.39(0.08) & 0.36(0.13) \\
   &                   &                     & 2.496 & 0.30(0.05) & 0.47(0.13) \\
R12 & 12:03:45.47(0.07) & $-$21:36:11.7(2.1) & 1.384 & 0.52(0.07) & 0.80(0.17) \\
   &                   &                     & 2.496 & ---        & --- \\
R13 & 12:03:47.24(0.05) & $-$21:36:15.1(1.3) & 1.384 & 0.78(0.07) & 1.01(0.16) \\
   &                   &                     & 2.496 & 0.31(0.05) & 1.09(0.19) \\
R14 & 12:03:47.91(0.04) & $-$21:28:33.1(0.5) & 1.384 & 5.53(0.08) & 5.32(0.14) \\
   &                   &                     & 2.496 & 2.70(0.06) & 2.79(0.12) \\[12pt]
\multicolumn{6}{c}{\normalsize \rxj} \\[2mm]

R1 & 13:14:00.90(0.04) & $-$25:16:53.7(0.7) & 1.384 & 2.09(0.09) & 2.38(0.1) \\
&                  &                        & 2.496 & 0.84(0.06) & 1.13(0.1) \\
R2 & 13:14:18.62(0.05) & $-$25:15:47.0(1.0) & 1.384 & 1.16(0.03) & 13.0 \\
&                  &                        & 2.496 & ---        & --- \\
R3 & 13:14:30.31(0.04) & $-$25:17:14.2(0.7) & 1.384 & 1.93(0.08) & 4.4 \\
&                  &                        & 2.496 & 0.54(0.04) & 1.7 \\
R4 & 13:14:34.31(0.06) & $-$25:11:59.9(1.6) & 1.384 & 0.69(0.09) & 0.64(0.2) \\
&                  &                        & 2.496 & 0.62(0.07) & 0.57(0.1) \\
R5 & 13:14:45.90(0.05) & $-$25:15:05.5(0.7) & 1.384 & 1.57(0.04) & 6.8 \\
&                  &                        & 2.496 & 0.48(0.04) & 1.5 \\
R6 & 13:14:48.64(0.04) & $-$25:16:18.3(0.6) & 1.384 & 3.20(0.09) & 3.65(0.2) \\
&                  &                        & 2.496 & 1.47(0.05) & 2.20(0.1) \\
    \hline
    \label{radiosrcs}
  \end{tabular}
  }
\end{table}

A search was carried out for optical identifications of the radio
sources in Table~\ref{radiosrcs} using the SuperCOSMOS catalogue
(Hambly et al. \cite{sss}).  A search radius of 10 arcsec was used.
The results are shown in Table~\ref{opticalid}.

\setlength{\tabcolsep}{0.2cm}

\begin{table}
\caption{Radio-optical identifications in \aco\ and \rxj. N refers to
  the identification number given in Table~\ref{radiosrcs} and
  $\Delta$r is the radius-vector offset between radio and SuperCOSMOS optical
  positions. T refers to the SuperCOSMOS image classification:
  1=galaxy; 2=star.}
{\footnotesize
\begin{tabular}{rccccc}
\hline
\hline
N & RA & Dec & $B_J$ & $\Delta$r & T \\
  & \multicolumn{2}{c}{Optical (J2000)} & (mag) & (\arcsec) \\
\hline \\[-2mm]
\multicolumn{6}{c}{\underline{\aco}}\\
R1  & 12:02:51.56  & $-$21:26:35.4 & 19.26 & 1.6 & 1\\
R2  & 12:02:56.81  & $-$21:28:55.8 & 18.42 & 9.1 & 1\\
R3  & 12:02:58.92  & $-$21:38:36.4 & 18.18 & 0.7 & 1\\
R5  & 12:03:08.68  & $-$21:39:40.0 & 20.12 & 0.7 & 2\\
R6  & 12:03:10.35  & $-$21:29:54.2 & 19.13 & 3.4 & 1\\
R7  & 12:03:17.47  & $-$21:32.27.4 & 19.41 & 4.2 & 1\\
R8  & 12:03:26.98  & $-$21:30:51.5 & 18.19 & 1.6 & 1\\
R11 & 12:03:45.70  & $-$21:36:11.9 & 18.98 & 3.5 & 1\\
R12 & 12:03:47.14  & $-$21:36:12.8 & 20.28 & 2.8 & 1\\
R13 & 12:03:48.11  & $-$21:28:33.1 & 20.40 & 3.2 & 2\\[3mm]
\multicolumn{6}{c}{\underline{\rxj}}\\
R1  & 13:14:00.89 & $-$25:16:54.3 & 19.81 & 0.6 & 2\\
R2  & 13:14:23.78 & $-$25:07:51.8 & 21.07 & 2.9 & 1\\
R3 & 13:14:30.36 & $-$25:17:17.4 & 21.40 & 3.3 & 1\\
R4 & 13:14:34.27 & $-$25:11:58.9 & 20.07 & 1.1 & 1\\
R5 & 13:14:46.17 & $-$25:15:09.1 & 21.37 & 5.1 & 1\\
\hline
\label{opticalid}
\end{tabular}
}
\end{table}

\subsubsection{\aco}

The 1.384 GHz (20 cm) radio contours overlaid on the DSS image of
the cluster field are shown in Fig.~\ref{fig13}\footnote{Available only
in the electronic version}; the 1.384 GHz (20 cm) radio contours for
the central part of the cluster are shown in Fig.~\ref{fig7} together
with the X-ray contours.  The central radio source (R7 in
Table~\ref{radiosrcs}) is extended north-south with a peak that best
matches galaxy \#40 from Table~\ref{tab3}.  The apparent wide-angle
tail (WAT) morphology relies on the lowest (2.5$\sigma$) contours and
is therefore uncertain.  Such morphologies are usually associated with
the central dominant cluster member.  An alternative interpretation,
suggested by the 13 cm observation shown only for the central part of
the cluster on Fig.~\ref{fig2}, is that R7 is a head-tail source
identified either with galaxy \#40 or the brighter galaxy \#39 which is
$\sim 10\arcsec$ north.  Higher resolution radio observations are
needed to settle this issue.

Apart from R7, there are no other cross-identifications between the
spectroscopic catalogue and radio source list. However, there are
several likely cluster identifications, including the head-tail source
(R1) located $\sim$8$\arcmin$ NW of the cluster centre.  The weakest
source in the field, R11, is a clear AGN candidate --- it is a strong
X-ray source with a flat radio spectrum; SuperCOSMOS classifies the
optical counterpart as a galaxy.  R9 is an extended source with no
obvious optical counterpart.  We initially considered it as a possible
relic source (En{\ss}lin et al.\ \cite{ens98}), but the 13 cm image
suggests an identification with a very faint object visible on both
the B and R sky survey images. The QSO (\#32, $z=1.17$) is radio
quiet.

\subsubsection{\rxj}

The 1.384 GHz (20 cm) radio contours overlaid on the DSS image of
the cluster field are shown in Fig.~\ref{fig14}\footnote{Available only
in the electronic version}; the central region of \rxj, together with
X-ray contours, are shown in Fig.~\ref{fig9} overlaid on the optical
image from Fig.~\ref{fig3}. Background noise in the central region of
the 20 cm image is affected by sidelobes of the 200 mJy source NVSS
J1314$-$2522 located SW of the cluster centre, just outside one Abell
radius.  For this reason the lowest contour level in Fig.~\ref{fig9}
has been set at $\sim$4$\sigma$.

As shown in Table \ref{radiosrcs}, the projected radio source density
is lower than for \aco.  There are no positive identifications with
spectroscopically confirmed cluster members, although there is a weak
source just below the 0.5 mJy threshold, very close to the brightest
cluster galaxy BCG2 (galaxy \#48; see Fig.\ \ref{fig9}).  There is
also a striking mirror symmetry of sources R2 and R5
(Table~\ref{radiosrcs}) with respect to this galaxy, resembling the
lobes of an FR II radio galaxy (see e.g. Fig.~\ref{fig9}). At the
redshift of the galaxy, $z=0.2466$, the linear size of $6\arcmin
15\arcsec$ would correspond to $\sim$1 Mpc, a typical scale for a
giant radio galaxy (Schoenmakers et al.\ \cite{scho01}). Such an
interpretation is questionable, however, as the galaxy is clearly not
located in an underdense environment (see Fig.~\ref{fig6}), as
required for the growth of giant sources. Neither source has an
obvious optical counterpart in the 20 cm image, but R5 appears double
at 13 cm, with a plausible optical identification midway between the
components.  On the other hand, with its extended radio emission,
steep spectrum and position in the cluster, R5 is an excellent
candidate for a relic source.  Its true extent is difficult to judge
from the ATCA 20 cm image because of sidelobe confusion, but there is
a suggestion of low surface brightness, extended emission in the
vicinity of BGC\#1 and the western X-ray peak.

\section{Discussion}
\label{sec:disc}

\subsection{\aco}

The dynamical state of \aco\ is very similar to that of Abell 665
(Gom\`ez et al.\ \cite{g00}) and to the more distant Abell 1300
(L\'emonon et al.\ \cite{ludo97}, Reid et al. \cite{rei98}),
suggesting that it may also be in the final stage of establishing
equilibrium after a merger event.

Support for the merger scenario comes from different morphological and
physical reasons which are summarized below:

\begin{itemize}
\item There is no single dominant galaxy. The brightest cluster galaxy
  (\#34, Table~\ref{tab3}) is $35\arcsec$ away from the X-ray centroid,
  and 580 \kms\ from the cluster mean redshift.\\
  The putative identification for the central radio source (\#40,
  Table~\ref{tab3}) is offset by 1300 \kms\ from the cluster mean
  redshift.
\item Nearly regular X-ray emission, without substructures but slightly
  twisted inner part.
\item No cooling flow region.
\item Deviations from the observed scaling relations $L_X$--$T$ and
$\sigma_v$--$T$ (Xue \& Wu \cite{xue00}), shown in
Table~\ref{tab:scaling}. The cluster is significantly less luminous
than expected from its measured temperature. The very high observed
temperature (13.4 keV) is probably an indication of a shock that
occurred in the recent past. We cannot exclude a possible
overestimation of the published ASCA temperature due to the presence
of the background X-ray point source. However, the small difference
(in the wrong sense) between the ASCA luminosity and our estimate with
the QSO emission excluded (see Table~\ref{tab:x}) seems unlikely to
account for a temperature overestimate of more than 4 keV. A new and
more accurate measurement of the temperature is clearly needed.
\item   Based on numerical simulations (Roettiger et al.\
  \cite{roe97}, Belsole et al.\ \cite{bel02}),
the regular X-ray morphology and high X-ray temperature
  are fully compatible with expectations from a past merger.
\end{itemize}

We can also determine the dynamical status of the cluster by comparing
its kinetic and potential energies.  From the measured velocity
dispersion we find $\beta_{\rm spec} = \mu m_p \sigma_v^2/kT =
0.84\pm0.25$, while from the X-ray emission, with the correction
factor from Bahcall \& Lubin (\cite{bah94}), we have $\beta^c_X = 1.25
\beta_X = 0.63\pm0.1$.  These values are consistent within the
uncertainties, indicating that the gas and galaxy motions are close to
equipartition.

If a merger occurred recently we might expect signatures at radio
wavelengths, such as radio halo/relic sources, and possibly tailed
sources (e.g.  En{\ss}lin et al.\ \cite{ens98}, Reid et al.\
\cite{rei98}, R\"ottgering et al.\ \cite{ro94}). There is no evidence
for a radio halo, although there is a tailed source (R7) near the
cluster centre which could have disrupted it (Giovannini \cite{gio99}, Liang
et al.\ \cite{lia00}).

\begin{table}
\caption{Scaling relations for \aco\ and \rxj. $\sigma_v$--$T$ is from
  Xue \& Wu (\cite{xue00}), while $L_{\rm bol}$--$T$ is from Arnaud \&
  Evrard (\cite{arn99}). The temperature from ASCA is in keV, $L_{\rm
  bol}$ in units of $10^{44}$ \ergs\ and $\sigma_v$ in \kms\ are from
  this paper.} \label{tab:scaling} {\footnotesize
\begin{tabular}{lrcccc}
\hline
\hline
Cluster & $T$~ & $\sigma_v$ & $\sigma_v$ & $L_{\rm bol}$ & $L_{\rm bol}$ \\
        & (obs) & (obs)      & ($\sigma_v$--$T$) & (obs)      & 
($L_{\rm bol}$--$T$)     \\
\hline \\[-2mm]
\aco    & 13.4 & 1330 & 1670 & 39.8 & 116  \\
\rxj    & 8.7  & 1100 & 1261 & 34.0 & 33.5 \\
\hline
\end{tabular}
}
\end{table}

\subsection{\rxj}

\rxj\ is morphologically and dynamically very different from \aco.  It
shows a clear bi-modal structure --- there are two groups in velocity
space separated by $\sim 1700$ \kms\ (cf. Table \ref{tab6} and
Fig.~\ref{fig5}) which are also separated in the projected galaxy
distribution (cf. Fig.~\ref{fig6}). The dominant galaxies of each group
are separated by $\sim 1000$ \kms\ in redshift space, and $2\arcmin
25\arcsec$, or $\approx$700 kpc, in projected distance. 

The X-ray emission is elongated, with the centroid located between the
two dominant galaxies. The elongation, however, is rotated by
$\approx$20$^{\circ}$ from the axis connecting the two BCGs.  This may
simply be due to the decoupling between the galaxies and gas during
the merger.

There are no cluster radio sources within the X-ray extension, with
the possible exception of the weak (uncatalogued) source at the
position of galaxy \#48.  If we are witnessing an interaction between
two sub-clusters, we might expect stronger radio activity than
observed.  However, residual sidelobes from a strong background source
$\sim 7\arcmin$ south of the centre hamper the detection of any very
extended emission. In addition, a more compact ATCA antenna
configuration is needed to improve sensitivity to low surface
brightness emission.  There are, however, two extended radio sources,
one of which (R2) has a steep radio spectrum and no optical
counterpart, and is therefore a plausible candidate for a relic
source.

The observed $L_X$, $T$ and $\sigma_v$ for \rxj\ are in good agreement
with the $L_X$--$T$ and $\sigma_v$--$T$ scaling relationships
(Table~\ref{tab:scaling}), suggesting that the merger has progressed
to the stage where the transient shock heating and radio activity have
dissipated.  On the other hand, if the cluster is in a pre-merging
phase, then it is unusual that the X-ray elongation is not aligned
with the group centres and that there is no sign of X-ray substructure
around the eastern group, as revealed, for example, in numerical
simulations (Roettiger et al. \cite{roe97}, Takizawa \cite{tak00}).
The scattered appearance of the projected galaxy distribution of the
eastern group compared to the western group (Fig.~\ref{fig6}) also
supports a post-merging scenario.

\bigskip

In conclusion, our observations suggest that neither cluster is
relaxed following a recent merger. However, their properties and
scaling laws are quite different, illustrating the diversity in
the merging and relaxation processes in cluster formation and
evolution.  The current data for the two clusters are compatible with
the expectations from the merger of a small group with a bigger
cluster for \aco, and nearly equal mass groups for \rxj.  Deep XMM and
Chandra observations, coupled with detailed numerical simulations
are needed to assess these hypotheses and better understand the many
aspects of the physical processes occurring during accretion and
relaxation over the course of a cluster merger.

\begin{acknowledgements}

  We would like to thank Romain Teyssier, John Hughes and Pierre-Alain
  Duc for numerous discussions about simulations, data reduction and
  analysis. We are especially indebted to Hector Flores and Dario
  Fadda for providing us with the observation of galaxy \#48 in \rxj\
  (CFHT, June 2001). We thank the referee Reinaldo de Carvalho for 
  valuable comments.

\end{acknowledgements}

\clearpage

\setcounter{table}{2}
\begin{table*}
\caption{A1451 spectroscopic catalogue. The columns are: reference
  number (``*'' denotes 1993 observations);  J2000 coordinates; redshift
  from cross-correlation or
  line measurements; the corresponding error; $B_J$ magnitude, or R when
  $B_J$ is not available; flag; and notes.
  Flag codes are:  0 -- star, 1 -- galaxy member of
  the cluster, 2 -- galaxy member of the cluster with emission lines,
  3 -- emission line galaxy not in the cluster, 5 -- other
  galaxy. Magnitudes are taken from the SuperCOSMOS Sky Survey
  (Hambly et al. \cite{sss}), ``$\dag$'' indicates possible magnitude problem
  due to blending.\label{tab3}}
{\scriptsize
\begin{tabular}{lccrclcl}
\hline\hline\\[-2mm] 
No. & R.A.\ (J2000) & Dec.\ (J2000) & $z$~~~ & $\Delta z$ &
~~$B_J$ & Flag & Notes\\ 
\hline\\[-2mm]
01  & 12:02:59.4 &  -21:33:41  & 0.0009 & 0.0003 & 17.81 &0 &\\
02  & 12:03:00.2 &  -21:35:01  & 0.1482 & 0.0008 & 19.16 &5 &\\
03  & 12:03:01.0 &  -21:32:44  & 0.1968 & 0.0007 & 21.82 &1 &\\
04  & 12:03:01.5 &  -21:33:47  & 0.0020 & 0.0005 & 19.01 &0 &\\
05  & 12:03:02.0 &  -21:34:20  & 0.1983 & 0.0008 & ---   &1 &\\
06  & 12:03:03.1 &  -21:34:08  & 0.2015 & 0.0004 & 20.78 &1 &\\
07  & 12:03:03.6 &  -21:32:25  & 0.3362 & 0.0010 & 21.68 &5 &\\
08  & 12:03:03.9 &  -21:32:02  & 0.2011 & 0.0007 & 21.35 &1 &\\
09  & 12:03:04.9 &  -21:32:52  & 0.1915 & 0.0004 & 19.58 &1 &\\
10  & 12:03:04.9 &  -21:33:17  & 0.2046 & 0.0009 & 20.71 &1 &\\
11  & 12:03:05.9 &  -21:32:36  & 0.1983 & 0.0009 & 22.35 &1 &\\
12  & 12:03:06.9 &  -21:32:11  & 0.0014 & 0.0004 & 18.34 &0 &\\
13  & 12:03:08.1 &  -21:33:54  & 0.2003 & 0.0006 & 21.29 &1 &\\
14  & 12:03:08.5 &  -21:32:08  & 0.2133 & 0.0007 & 19.19 &5 &\\
15  & 12:03:09.1 &  -21:33:20  & 0.1928 & 0.0008 & 21.20 &1 &\\
16  & 12:03:09.8 &  -21:33:46  & 0.2006 & 0.0004 & 20.36 &1 &\\
17  & 12:03:10.4 &  -21:33:56  & 0.0007 & 0.0004 & 20.10 &0 &\\
18  & 12:03:10.2 &  -21:32:45  & 0.3071 & 0.0005 & ---   &5 &\\
19* & 12:03:11.3 &  -21:30:24  & 0.1985 & 0.0007 & 21.64 &1 & \\
20  & 12:03:12.1 &  -21:34:15  & 0.2049 & 0.0004 & 20.62 &1 &\\
21  & 12:03:12.9 &  -21:32:58  & 0.3053 & 0.0003 & 22.07 &3 &[OII],[OIII]\\
22  & 12:03:13.2 &  -21:33:35  & 0.1879 & 0.0008 & 22.72 &1 &\\
23  & 12:03:13.8 &  -21:33:23  & 0.0018 & 0.0005 & 21.35 &0 &\\
24  & 12:03:14.2 &  -21:34:04  & 0.2066 & 0.0006 & 18.99 &2 &[OII],H$_\beta$,[OIII],H$_\alpha$\\
25* & 12:03:14.5 &  -21:29:50  & 0.2032 & 0.0004 & 20.08$^\dag$ &1 &\\
26* & 12:03:14.5 &  -21:31:31  & 0.2037 & 0.0007 & 19.91 &1 &\\
27  & 12:03:14.5 &  -21:33:49  & 0.2044 & 0.0004 & 19.96 &2 &[OII],[OIII],H$_\alpha$\\
28  & 12:03:14.9 &  -21:32:05  & 0.0019 & 0.0005 & 19.61 &0 &\\
29* & 12:03:15.2 &  -21:30:23  & 0.1909 & 0.0004 & 19.37 &1 &\\
30  & 12:03:15.4 &  -21:34:25  & 0.0009 & 0.0007 & 17.39 &0 &\\
31  & 12:03:15.7 &  -21:34:09  & 0.2037 & 0.0003 & 21.05 &1 &\\
32  & 12:03:15.9 &  -21:33:43  & 1.1710 & 0.001 &  20.33 &3 &QSO: CIII],MgII
\\
33  & 12:03:16.4 &  -21:32:57  & 0.1946 & 0.0004 & ---   &1 &\\
34  & 12:03:16.7 &  -21:32:54  & 0.1966 & 0.0005 & 18.53 &1 &\\
35  & 12:03:16.8 &  -21:32:47  & 0.2986 & 0.0007 & 20.76 &5 &\\
36* & 12:03:16.9 &  -21:31:59  & 0.1989 & 0.0006 & 19.19$^\dag$  &1 &\\
37  & 12:03:17.1 &  -21:34:16  & 0.2133 & 0.0003 & 19.82 &3 &\\
38* & 12:03:17.3 &  -21:32:21  & 0.1948 & 0.0006 & ---   &1 &\\
39* & 12:03:17.4 &  -21:32:19  & 0.1995 & 0.0004 & 18.78$^\dag$  &1 &\\
40* & 12:03:17.5 &  -21:32:28  & 0.2041 & 0.0006 & 19.41 &1 &\\
41* & 12:03:17.5 &  -21:31:29  & 0.1994 & 0.0005 & 20.44 &1 &\\
42  & 12:03:17.8 &  -21:33:40  & 0.3025 & 0.0007 & 20.60 &3 &[OII],[OIII]\\
43  & 12:03:17.8 &  -21:33:21  & 0.1923 & 0.0005 & 20.02 &1 &\\
44  & 12:03:18.1 &  -21:34:55  & 0.1952 & 0.0005 & 19.68 &1 &\\
45  & 12:03:18.5 &  -21:34:45  & 0.1931 & 0.0005 & 21.19(R) &1 &\\
46  & 12:03:18.7 &  -21:34:10  & 0.0010 & 0.0006 & 20.90 &0 &\\
47  & 12:03:18.8 &  -21:33:54  & 0.2003 & 0.0003 & 19.70 &1 &\\
48  & 12:03:19.0 &  -21:33:20  & 0.3341 & 0.0004 & 20.79 &5 &\\
49* & 12:03:19.2 &  -21:29:38  & 0.1989 & 0.0006 & 20.58 &1 &\\
50* & 12:03:19.2 &  -21:31:24  & 0.2085 & 0.0007 & 20.37 &1 &\\
51* & 12:03:19.2 &  -21:30:49  & 0.1941 & 0.0008 & 20.12 &1 &\\
52  & 12:03:19.3 &  -21:33:27  & 0.1965 & 0.0006 & 22.22 &1 &\\
53  & 12:03:19.3 &  -21:33:20  & 0.1897 & 0.0004 & 19.02 &1 &\\
54* & 12:03:19.6 &  -21:30:02  & 0.1931 & 0.0005 & 19.73 &1 &\\
55  & 12:03:19.6 &  -21:33:04  & 0.1971 & 0.0007 & 20.69 &1 &\\
56  & 12:03:20.2 &  -21:32:49  & 0.1986 & 0.0004 & 21.13 &1 &\\
57  & 12:03:20.5 &  -21:33:51  & 0.3353 & 0.0004 & 20.92 &5 &\\
58  & 12:03:20.8 &  -21:33:06  & 0.2004 & 0.0006 & 21.60 &1 &\\
59  & 12:03:21.0 &  -21:33:10  & 0.1924 & 0.0005 & 20.31 &1 &\\
60* & 12:03:21.1 &  -21:32:41  & 0.2092 & 0.0009 & 20.19 &1 &\\
61  & 12:03:21.2 &  -21:32:38  & 0.2012 & 0.0005 & 20.92 &1 &\\
62  & 12:03:21.3 &  -21:35:03  & 0.0001 & 0.0008 & 16.28 &0 &\\
63  & 12:03:22.2 &  -21:34:12  & 0.1897 & 0.0003 & 20.75 &1 &\\
64  & 12:03:22.2 &  -21:33:23  & 0.2049 & 0.0003 & 19.16 &1 &\\
65* & 12:03:22.5 &  -21:29:57  & 0.1943 & 0.0006 & 20.64 &1 &\\
66  & 12:03:23.5 &  -21:32:58  & 0.1929 & 0.0008 & 21.23 &1 &\\
67  & 12:03:23.7 &  -21:32:48  & 0.2021 & 0.0004 & 19.81 &1 &\\
68  & 12:03:24.8 &  -21:35:06  & 0.1985 & 0.0006 & 20.39 &1 &\\
69  & 12:03:25.2 &  -21:32:59  &-0.0011 & 0.0003 & 16.84 &0 &\\
70  & 12:03:25.7 &  -21:32:24  & 0.2043 & 0.0007 & 20.27 &1 &\\
71  & 12:03:26.2 &  -21:33:33  & 0.2000 & 0.0003 & 19.77 &2 &[OII],[NeIII],H$_\beta$,[OIII],H$_\alpha$\\
72  & 12:03:27.0 &  -21:31:54  & 0.2045 & 0.0007 & 20.13 &1 &\\
73  & 12:03:27.0 &  -21:32:35  & 0.2044 & 0.0006 & 21.86 &1 &\\
74  & 12:03:27.3 &  -21:33:58  & 0.2011 & 0.0004 & 20.89 &1 &\\
75  & 12:03:29.3 &  -21:33:14  & 0.5167 & 0.0005 & 19.41(R) &5 &\\
76  & 12:03:30.8 &  -21:31:55  &-0.0001 & 0.0003 & 15.69 &0 &\\
77  & 12:03:31.2 &  -21:32:12  & 0.1968 & 0.0003 & 18.99 &2 &[OII],[NeIII],H$_\gamma$,H$_\beta$\\
78  & 12:03:31.5 &  -21:32:36  & 0.2093 & 0.0010 & 21.57 &1 &\\
79  & 12:03:32.4 &  -21:33:51  & 0.2025 & 0.0004 & 20.52 &1 &\\
80  & 12:03:33.0 &  -21:33:01  & 0.1946 & 0.0004 & 21.29 &1 &\\
81  & 12:03:33.6 &  -21:32:44  & 0.0017 & 0.0003 & 15.38 &0 &\\
\hline
\end{tabular}
}
\end{table*}

\setcounter{table}{3}
\begin{table*}
\caption{\rxj\ spectroscopic catalogue; see caption for
  Table~\ref{tab3}.\label{tab4} } {\scriptsize
\begin{tabular}{lccrclcl}
\hline\hline\\[-2mm]
No. & R.A.\ (J2000)& Dec.\ (J2000)   & $z$~~~    & $\Delta z$ & ~~$B_J$ &
Flag & Notes\\
\hline\\[-2mm]
01  &  13:14:14.7 &  -25:13:11 &   0.0020 &  0.0007& 19.10  &0 & \\
02  &  13:14:17.3 &  -25:13:48 &   0.2529 &  0.0009& 20.71  &1 & \\
03  &  13:14:17.8 &  -25:17:09 &   0.2537 &  0.0005& 19.78  &1 & \\
04  &  13:14:17.7 &  -25:15:37 &   0.2394 &  0.0008& 21.66  &1 & \\
05  &  13:14:18.1 &  -25:20:18 &  -0.0005 &  0.0005& 19.67  &0 & \\
06  &  13:14:19.2 &  -25:14:15 &   0.2496 &  0.0005& 20.07  &1 & \\
07  &  13:14:19.3 &  -25:15:30 &   0.2501 &  0.0005& 20.17  &1 & \\
08  &  13:14:19.5 &  -25:13:37 &   0.2520 &  0.0007& 17.76(R)&1 & \\
09  &  13:14:19.6 &  -25:15:38 &   0.0001 &  0.0003& 16.19  &0 & \\
10  &  13:14:20.0 &  -25:17:00 &   0.2491 &  0.0008& 22.30  &1 & \\
11  &  13:14:20.3 &  -25:15:05 &   0.2555 &  0.0007& 20.92  &1 & \\
12  &  13:14:20.9 &  -25:17:36 &   0.2533 &  0.0005& 20.83  &1 & \\
13  &  13:14:21.0 &  -25:16:58 &  -0.0014 &  0.0003& 17.25  &0 & \\
14  &  13:14:21.0 &  -25:16:11 &   0.2477 &  0.0006& 22.22  &1 & \\
15  &  13:14:21.4 &  -25:17:36 &   0.2939 &  0.0002& 21.65  &5 & $z$ from lines\\
16  &  13:14:21.6 &  -25:16:31 &   0.2493 &  0.0005& 21.31  &1 & \\
17  &  13:14:21.8 &  -25:15:23 &   0.2538 &  0.0008& 21.65  &1 & \\
18  &  13:14:21.9 &  -25:16:22 &   0.2504 &  0.0004& 19.27  &1 & \\
19  &  13:14:22.1 &  -25:15:46 &   0.2503 &  0.0004& 18.43  &1 & BCG \#1\\
20  &  13:14:22.1 &  -25:15:45 &   0.2487 &  0.0004& ---    &1 & \\
21  &  13:14:22.1 &  -25:14:34 &  -0.0012 &  0.0003& 15.57  &0 & \\
22  &  13:14:22.4 &  -25:17:19 &   0.0008 &  0.0003& 15.94  &0 & \\
23* &  13:14:23.1 &  -25:16:51 &   0.2443 &  0.0004& 20.85  &1 & \\
24  &  13:14:23.3 &  -25:12:35 &   0.2453 &  0.0006& 20.99  &1 & \\
25  &  13:14:23.6 &  -25:16:40 &   0.2503 &  0.0005& 21.35  &1 & \\
26  &  13:14:24.1 &  -25:17:09 &   0.2485 &  0.0006& 21.51  &1 & \\
27  &  13:14:24.2 &  -25:14:45 &   0.2466 &  0.0005& 21.20  &1 & \\
28  &  13:14:24.2 &  -25:13:56 &   0.2492 &  0.0004& 20.23  &1 & \\
29  &  13:14:24.4 &  -25:16:06 &  -0.0010 &  0.0003& 15.33  &0 & \\
30  &  13:14:24.5 &  -25:16:36 &   0.2453 &  0.0006& 21.44  &1 & \\
31  &  13:14:24.7 &  -25:13:42 &   0.2424 &  0.0005& ---    &1 & \\
32  &  13:14:24.7 &  -25:13:38 &   0.2376 &  0.0005& ---    &1 & \\
33* &  13:14:24.9 &  -25:15:09 &   0.2494 &  0.0007& 20.80  &1 & \\
34  &  13:14:25.9 &  -25:14:33 &   0.2478 &  0.0005& 21.63  &1 & \\
35  &  13:14:26.1 &  -25:17:01 &   0.2491 &  0.0006& 21.74  &1 & \\
36* &  13:14:26.2 &  -25:15:22 &   0.0008 &  0.0009& 21.08  &0 & $z$ from lines \\
37  &  13:14:26.4 &  -25:16:49 &   0.2426 &  0.0006& 21.53  &1 & \\
38  &  13:14:27.3 &  -25:18:41 &   0.2449 &  0.0007& 22.11  &1 & \\
39* &  13:14:27.8 &  -25:15:40 &   0.2429 &  0.0005& 20.87  &1 & \\
40* &  13:14:28.4 &  -25:16:31 &   0.2435 &  0.0004& 20.47  &1 & \\
41  &  13:14:28.6 &  -25:14:52 &   0.2510 &  0.0006& ---    &1 & \\
42* &  13:14:29.2 &  -25:16:13 &   0.2425 &  0.0005& 20.60  &1 & \\
43  &  13:14:30.0 &  -25:15:58 &   0.0000 &  0.0007& 15.52  &0 & \\
44* &  13:14:33.8 &  -25:15:48 &   0.2411 &  0.0006& 21.48  &1 & \\
45* &  13:14:34.5 &  -25:15:31 &   0.2399 &  0.0007& 20.91  &1 & \\
46* &  13:14:35.6 &  -25:15:46 &  -0.0001 &  0.0007& 22.24  &0 & \\
47* &  13:14:38.0 &  -25:17:06 &   0.2433 &  0.0002& 19.33  &2 &Sy1: 
[OII],[NeIII],H$_\beta$,[OIII] \\
48 &   13:14:32.9 &  -25:15:28 &   0.2463 &  0.0004& 18.30 &1 & BCG \#2\\
\hline
\end{tabular}
}
\end{table*}

\end{document}